\documentclass[11pt]{article}

\usepackage{graphicx}
\begin{document}

\begin{center}
\mbox{}\vskip-3cm\mbox{}
\footnotesize\it
Published in: 
J.~Ashkenazi et al.~(eds.), New Challenges in Superconductivity: Experimental Advances and Emerging Theories,  pp.~85--90 (Springer, New York, 2005)\\ \mbox{}\\
\hrule\mbox{}\vskip2cm\mbox{}
\end{center}

\begin{center}
{\LARGE\bf Measurements of the doping effects\\ on the in-plane paraconductivity\\ in cuprate superconductors\\}
\mbox{}\\
\mbox{}\\
F\'elix Vidal, Manuel V. Ramallo, Gonzalo Ferro, Jos\'e Antonio Veira\\
\mbox{}\\
{\it
Laboratorio de Baixas Temperaturas e Superconductividade\\ (Unidad Asociada al ICMM, CSIC, Madrid, Spain)\\ Departamento de F\'{\i}sica da Materia Condensada,\\ Universidade de Santiago de Compostela,\\ Santiago de Compostela E15782 Spain
}
\mbox{}\\
\mbox{}\\
\end{center}

 \begin{abstract}
 We will summarize here some of our measurements of the superconducting fluctuations effects on the in-plane electrical resistivity (the so-called  in-plane paraconductivity) in La$_{2-x}$Sr$_x$CuO$_4$ thin films with different Sr content. Our results suggest that these superconducting fluctuations effects are not related to the opening of a pseudogap in the normal-state of underdoped compounds.
 \end{abstract}

\section{Introduction: the questions that are and aren't addressed here}
In Fig.~1, we represent an example of the temperature behaviour of the electrical resistivity of underdoped La$_{1.9}$Sr$_{0.1}$CuO$_4$. These data were taken from Refs.~\cite{BatloggPhaC} and \cite{CurrasPRB}, and correspond to a bulk polycrystalline sample (Fig.~1a) and to a thin film (Fig.~1b). In these figures, $T_C$ is the temperature where $\mbox{d}\rho/\mbox{d}T$ has its maximum ($T_C$ is close to $T_{C0}$, the temperature where the measured resistivity vanishes), $T^*$ is the temperature at which the pseudogap in the normal state opens, and $T^C$ is the temperature at which the measured resistivity, $\rho(T)$, becomes indistinguishable from the background resistivity, $\rho_B(T)$. This background resistivity (solid line in Fig.~1b) is obtained by extrapolating through the transition the normal-state resistivity measured well above the superconducting transition. The details of this extrapolation procedure may be seen in Ref.~\cite{CurrasPRB}.

\begin{figure}[t]
\begin{center}
\includegraphics[scale=0.5]{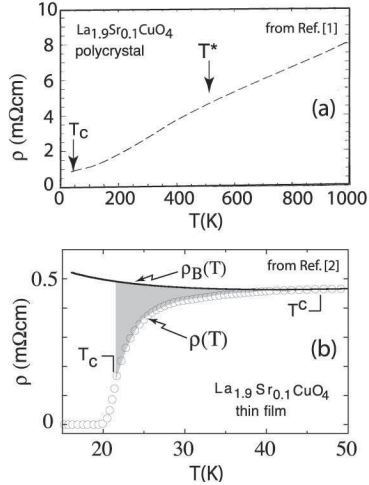}\mbox{}\vskip-2.5em\mbox{}\\
\end{center}
\caption{\sl Temperature dependence of the resistivity of (a) a polycrystalline sample of the underdoped cuprate La$_{1.9}$Sr$_{0.1}$CuO$_4$, taken from \cite{BatloggPhaC}, and (b) a film of the same composition and thickness $\sim150$~nm, taken from \cite{CurrasPRB}. In (b), the measurements correspond to the in-plane direction (parallel to the CuO$_2$ layers). The temperatures $T^*$, $T^C$ and $T_C$ correspond, respectively, to the pseudogap opening temperature, to the temperature where the SCF effects become indistinguishable, and to the (inflexion-point) observed normal-superconducting transition temperature.}
\mbox{}\vskip-2.5em\mbox{}\\
\end{figure}

The difference between $\rho(T)$ and $\rho_B(T)$ (dark region in Fig.~1b) is supposed to be due to the presence of coherent Cooper pairs created in the normal state by thermal fluctuations. As usual, these superconducting fluctuations (SCF) effects may be quantified through the so-called in-plane paraconductivity, defined by
%%%
\begin{equation}
\Delta\sigma(T)\equiv\frac{1}{\rho(T)}-\frac{1}{\rho_B(T)}.
\end{equation}
%%%

The two questions that we will address here are: {\it i)} How $\Delta\sigma(T)$ is affected by doping? This question may be directly answered by the experiments, independently of any theoretical approach. {\it ii)} Is the observed $T_C$ a ``good'' mean-field critical temperature for the SCF, even up to $T^C$? This second question is related to the theoretical description of the SCF. We will answer it on the grounds of the Gaussian-Ginzburg-Landau (GGL) approach, extended up to $T^C$ by introducing a ``total-energy'' cutoff that takes into account the limits imposed  by the uncertainty principle to the shrinkage of the superconducting wave function when the temperature increases well above $T_C$.\cite{CurrasPRB,VidalEPL} We are not going to directly address here another important (mainly to discriminate between the existing theoretical proposals for the pseudogap\cite{ReviewPseudogap}) open question: Is $T^*$ the ``true'' mean-field superconducting transition temperature? 

To answer the above questions we will present measurements of the in-plane paraconductivity in La$_{2-x}$Sr$_x$CuO$_4$ thin films with different Sr concentrations. Our experiments were detailed elsewhere.\cite{CurrasPRB}. They extend to high reduced-temperatures the earlier measurements of Hikita and Suzuki.\cite{HS} Other original aspect of our present work is the analysis in terms of the ``extended'' GGL approach (other analyses by other groups of the paraconductivity at high reduced-temperatures in different HTSC\cite{Silva,Leridon} did not take into account the quantum effects on the SCF, see also later).

\section{How are affected by doping the superconducting fluctuations above $T_C$?}
Some examples of the influence of the Sr content on the in-plane paraconductivity curves, $\Delta\sigma(\epsilon)_x$, measured as a function of the reduced temperature, $\epsilon\equiv\ln(T/T_C)$, are shown in Fig.~2. These data correspond to La$_{2-x}$Sr$_x$CuO$_4$ thin films, of about 150~nm thickness and grown on (100)SrTiO$_3$ substrates. Other experimental details may be seen in Ref.\cite{CurrasPRB}

The ``as-measured'' data of Fig.~2 provide a direct answer to the question stated in the title of this section: The only appreciable effects of doping on the SCF is to change somewhat the slope of the $\log\Delta\sigma(\epsilon)_x$ versus $\log\epsilon$ curves in the low reduced-temperature region. We will see in the next section that these changes are associated with changes in the SCF dimensionality. However, these changes manifest themselves only in the overdoped regime: As directly illustrated by the data in Fig.~2a, in all the temperature range of our measurements (approximately $10^{-2}\leq\epsilon\leq1$), the data for $x=0.1$ and $x=0.12$, which are well in the underdoped regime, agree well within the experimental uncertainties with those of the optimally-doped film ($x=0.15$). Moreover, the results of Figs.~2a to 2c show that in all the cases the SCF vanish above a reduced temperature of around 0.7, which corresponds to $T^C\sim2T_C$. Note already here that in the underdoped compounds $T^C\ll T^*$ (see also section \ref{seccioncuatro}, and Figs.~1 and 3). These different results already suggest, therefore, that the opening of the normal-state pseudogap is not related to the coherent Cooper pairs created above the superconducting transition by thermal fluctuations. 

\begin{figure}[t]
\begin{center}
\includegraphics[scale=0.60]{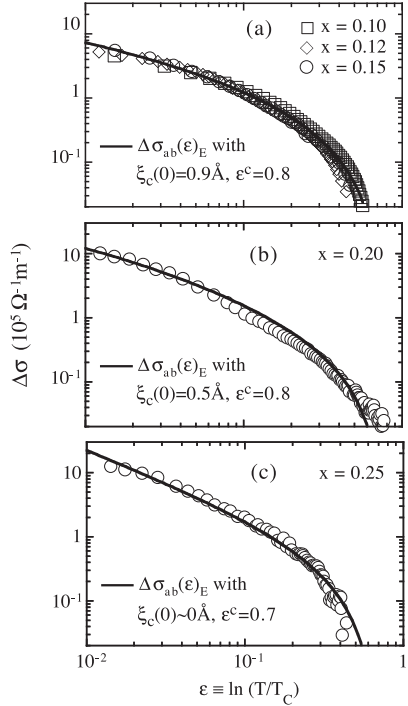}\mbox{}\vskip-3em\mbox{}\\
\end{center}
\caption{\sl In-plane paraconductivity versus reduced temperature measured in Ref.\cite{CurrasPRB}, in La$_{2-x}$Sr$_x$CuO$_4$ films with thickness $\sim150$~nm and Sr contents covering the underdoped ($x<0.15$), optimally-doped ($x=0.15$) and overdoped ($x>0.15$) compositions. The solid lines correspond to the best fit to those data using the GGL approach with a total-energy cutoff (Eq.~(\ref{teor})). See main text for details.}
\mbox{}\vskip-2.5em\mbox{}\\
\end{figure}

\section{Is the observed $T_C(x)$ a ``good'' mean--field super\-conducting--normal transition temperature?}
This question is also answered in Fig.~2, where we compare the experimental paraconductivity with the GGL approach extended to high reduced-temperatures by introducing a ``total-energy'' cutoff. This cutoff takes into account the limits imposed by the uncertainty principle to the shrinkage of the superconducting wave function, associated to the coherent Cooper pairs created by thermal fluctuations, when the temperature increases well above $T_C$.\cite{VidalEPL} The paraconductivity was calculated on the grounds of this ``extended'' GGL approach by Carballeira et al.\cite{CarballeiraDs} The corresponding expressions for a single-layered superconductor, which is the case well adapted to La$_{2-x}$Sr$_x$CuO$_4$, is\cite{CurrasPRB,CarballeiraDs}
%%%%
\begin{equation}
\Delta\sigma(\epsilon)_{\rm E}={{\mbox{e}^2}\over{16\hbar
s}}
\left[
{{1}\over{\epsilon}}
\left(
1+{{B_{\rm LD}}\over{\epsilon}}
\right)^{-1/2}
-
{{1}\over{\epsilon^C}}\left(2-
{{\epsilon+B_{\rm LD}/2}\over{\epsilon^C}}
\right)\right],\label{teor}
\end{equation}
%%%%
where $\hbar$ is the reduced Planck constant, e is the electron charge, $s$ is the superconducting CuO$_2$ layers periodicity ($s=0.66$~nm for La$_{2-x}$Sr$_x$CuO$_4$), $B_{\rm LD}\equiv[2\xi_c(0)/s]^2$ is the Lawrence-Doniach (LD) parameter and $\xi_c(0)$ is the superconducting coherence length amplitude in the $c$-direction (perpendicular to the CuO$_2$ layers). Note that in calculating this expression for $\Delta\sigma(\epsilon)_{\rm E}$ we have assumed the BCS-like value for the SCF relaxation time, $\tau_0=(\pi\hbar/8k_BT_C)\epsilon^{-1}$ (for more details see Refs.\cite{CurrasPRB,CarballeiraDs,tau}). The solid lines in Fig.~2 are the best fits of Eq.~(\ref{teor}) to the data points, with $\xi_c(0)$ and $\epsilon^C$, the reduced-temperature where $\Delta\sigma(\epsilon)_x$ vanishes, as the only free parameters.

\begin{figure}[tb]
\begin{center}
\includegraphics[scale=0.45]{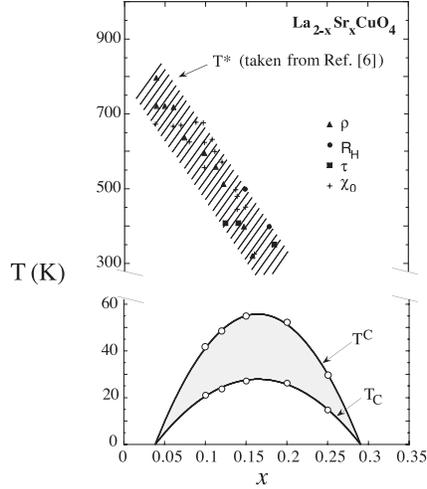}\mbox{}\vskip-2.5em\mbox{}\\
\end{center}
\caption{\sl Comparison for La$_{2-x}$Sr$_x$CuO$_4$ between the temperatures of the pseudogap opening ($T^*$), of the vanishing of the in-plane paraconductivity ($T^C$), and of the normal-superconducting transition as observed in the resistivity measurements ($T_C$). The $T^*$ values are taken from the compilation in Ref.~\cite{TPseudogap}, and correspond to resistivity measurements ($\rho$), Hall coefficient measurements ($R_H$), infrared measurements ($\tau$), and  static susceptibility measurements ($\chi_0$)}
\mbox{}\vskip-2.5em\mbox{}\\
\end{figure}

As may be observed in these figures, for all the samples the agreement between Eq.~(\ref{teor}) and the experimental data is excellent, in the entire studied $\epsilon$-region. The central result here is that this comparison strongly suggests then that the observed $T_C(x)$ is a ``good'' mean-field transition temperature for the measured SCF. This comparison also confirms the existence of a well-defined reduced-temperature, $\epsilon^C$, of around 0.7, above which the in-plane paraconductivity vanishes. Taking into account the experimental uncertainties,\cite{CurrasPRB} this value matches fairly well the value $\sim0.6$ which may be roughly estimated on the grounds of the uncertainty principle, the mean-field $\epsilon$-dependence of the in-plane coherence length, and the BCS relationship between the Pippard and the Ginzburg-Landau coherence lengths.\cite{CurrasPRB,VidalEPL} Moreover, by comparing the obtained values of $\xi_c(0)$, which are given in the corresponding figures, with $s$, it is easy to conclude that in the underdoped and optimally doped samples the SCF have a 2D-3D crossover around $\epsilon\simeq7\times10^{-2}$. In contrast, the SCF in the overdoped compounds are 2D in almost all the accessible $\epsilon$-region.

\section{Conclusions\label{seccioncuatro}}
The main conclusions of the experimental results and analyses presented here may be summarized as follows:

{\bf A model-independent conclusion: }
The general behaviour of the in-plane paraconductivity is {\em not} affected, even up to $T^C$, by doping. The SCF effects in  La$_{2-x}$Sr$_x$CuO$_4$ thin films seem to be not related to the pseudogap. 

{\bf From the comparison of the measured paraconductivity with the ``extended'' GGL approach: } {\it i)} The measured $T_C$ is a good mean-field critical temperature for the GGL approach. {\it ii)} Both the relaxation time of the SCF and the reduced temperature, $\epsilon^C$, where the SCF vanish, are doping-independent and they take values close to those of BCS superconductors. This last result demands further studies.

The independence of the SCF and the pseudogap seems to be confirmed when three characteristic temperatures, $T_C$, $T^C$ and $T^*$, are compared. This is done in Fig.~3, where the doping dependence of $T_C$, $T^C$ and $T^*$ is represented. The data for $T^*$ where taken from Ref.~\cite{TPseudogap}. This figure illustrates that in  the underdoped La$_{2-x}$Sr$_x$CuO$_4$ superconductors not only $T_C$ but also $T^C$ is much lower than $T^*$ and that the doping behaviour of both $T_C$ and $T^C$ is very different from the one of $T^*$.

\end{document}